\documentclass[onecolomn, journal]{IEEEtran}

\usepackage{multirow}
\usepackage{amsmath}
\usepackage{graphicx}
\usepackage{epstopdf}
\usepackage{bm}
\usepackage{amsfonts}
\usepackage[colorlinks,linkcolor=black,anchorcolor=black,citecolor=black]{hyperref}
\usepackage{cite}
\usepackage{makecell}
\usepackage{booktabs}
\usepackage{amssymb}
\usepackage{tabularx}
\usepackage{float}
\usepackage{subcaption}

\makeatletter
\newif\if@restonecol
\makeatother

\usepackage[ruled,vlined]{algorithm2e}
\usepackage{algpseudocode}
\usepackage{amsmath}
\mathchardef\mhyphen="2D

\newcolumntype{I}{!{\vrule width 3pt}}
\newlength\savedwidth

\newlength\savewidth

\begin{document}

\title{Csi-LLM: A Novel Downlink Channel Prediction Method Aligned with LLM Pre-Training
}

\author{
        Shilong~Fan, Zhenyu~Liu, Xinyu~Gu, Haozhen~Li


\thanks{S. Fan, H. Li, X. Gu are with the School of Artificial Intelligence, Beijing University of Posts and Telecommunications (BUPT), Beijing 100876, China. (e-mail: \{fansl, lihaozhen, guxinyu\}@bupt.edu.cn). Z. Liu is with the 5GIC and 6GIC, Institute for Communication Systems, University of Surrey, United Kingdom (e-mail: zhenyu.liu@surrey.ac.uk).
 X. Gu is also with the Purple Mountain Laboratories, Nanjing 211111, China. 
}
}

\maketitle

\begin{abstract}
Downlink channel temporal prediction is a critical technology in massive multiple-input multiple-output (MIMO) systems. However, existing methods that rely on fixed-step historical sequences significantly limit the accuracy, practicality, and scalability of channel prediction. Recent advances have shown that large language models (LLMs) exhibit strong pattern recognition and reasoning abilities over complex sequences. The challenge lies in effectively aligning wireless communication data with the modalities used in natural language processing to fully harness these capabilities. In this work, we introduce Csi-LLM, a novel LLM-powered downlink channel prediction technique that models variable-step historical sequences. To ensure effective cross-modality application, we align the design and training of Csi-LLM with the processing of natural language tasks, leveraging the LLM's next-token generation capability for predicting the next step in channel state information (CSI). Simulation results demonstrate the effectiveness of this alignment strategy, with Csi-LLM consistently delivering stable performance improvements across various scenarios and showing significant potential in continuous multi-step prediction.
\end{abstract}


\IEEEpeerreviewmaketitle

\section{Introduction}

\IEEEPARstart{M}{assive} multiple-input multiple-output (MIMO) is a pivotal technology in wireless communication, greatly enhancing spectral efficiency. The performance of MIMO systems, however, is closely tied to the effectiveness of precoding, which in turn depends on the quality of the estimated instantaneous channel state information (CSI) \cite{truong2013effects}. Unfortunately, in frequency division duplex (FDD) mode, feedback delays, and in time division duplex (TDD) mode, processing delays, can result in outdated CSI. Additionally, user mobility, particularly in high-mobility scenarios, further degrades the accuracy of CSI estimation. To mitigate the performance loss due to channel aging, extensive research has focused on channel prediction, leveraging the temporal correlation between historical and future channel states\cite{duel2007fading}.

Existing temporal correlation-based channel prediction methods can be broadly categorized into statistical-based approaches\cite{baddour2005autoregressive,yin2020addressing} and neural network-based techniques\cite{kim2020massive,jiang2019neural,jiang2022accurate}. Statistical prediction methods, such as those based on the auto-regressive (AR) model\cite{baddour2005autoregressive} and Prony’s method\cite{yin2020addressing}, rely on constructing mathematical models and applying historical data to these models. However, due to simplified channel assumptions and rapid channel variations, these traditional statistical methods often fail to provide accurate and real-time CSI predictions.

To overcome these limitations, recent research has shifted towards neural network-based downlink channel prediction methods, which offer a data-driven approach to more accurately model actual channels. For instance, \cite{kim2020massive} utilized fully-connected neural networks (FCNs) to predict future channels by learning from historical channel data. To address the high-dimensionality of input channels, \cite{jiang2019neural} employed recurrent neural networks (RNNs) to process historical channels iteratively in chronological order. To mitigate the rapid accumulation of prediction errors inherent in sequential channel prediction, \cite{jiang2022accurate} introduced a parallel prediction method that forecasts future channels over several steps simultaneously.

However, these methods typically rely on fixed-length historical sequences as inputs, which imposes significant constraints. The rigid structure of input and output reduces model scalability and practicality, and the challenges associated with data collection limit the potential for further enhancement. Note that, channel coherence time varies inversely with carrier frequency and user speed, meaning that the length of the correlated historical sequence can fluctuate with user movement, thereby degrading model performance. Moreover, customizing a separate model for each channel condition is often inefficient, leading to increased learning and hardware costs, including memory usage and model-switching overhead. 

Large Language Models (LLMs) like ChatGPT, GPT-4 (OpenAI, 2023)\cite{openai2024gpt4technicalreport}, and PaLM\cite{JMLR:v24:22-1144} represent significant milestones in AI development. Recently, researchers in the wireless communication field have begun investigating how to harness the unique capabilities of LLMs to optimize downstream tasks\cite{tong2023ten}. However, the training processes and data formats for LLMs, designed primarily for natural language tasks, differ substantially from those of traditional neural networks used in wireless communication. As a result, directly transferring models and converting modalities to apply LLMs in this domain may not fully exploit their powerful capabilities.

To design a flexible model that adapts to varying mobility scenarios while exploring the potential of LLMs in wireless communication, this letter introduces a novel LLM-based channel prediction model. This model addresses the challenge of variable input lengths by optimizing the training process and data format to align with the capabilities of LLMs. Our key contributions are as follows:

\begin{itemize}
\item We propose a novel LLM-aligned optimization strategy for downlink channel prediction neural networks. This strategy aligns network architecture, data processing, and optimization objectives with those of LLMs, maximizing their pattern recognition and reasoning abilities in wireless communication scenarios.
\item Building on this alignment strategy, we develop a new LLM-empowered downlink channel prediction scheme, termed Csi-LLM. Unlike traditional models that rely on fixed-length historical sequences, Csi-LLM supports variable-length inputs, enabling continuous and accurate downlink channel prediction.
\end{itemize}

\section{System Model and CSI Prediction}
\subsection{Massive MIMO OFDM Systems}
Consider a massive MIMO gNB equipped with \( N_t \) antennas, serving a number of UEs with \( N_r \gg 1 \) receiver antennas within its coverage area\cite{9839111}. Orthogonal frequency division multiplexing (OFDM) is employed for downlink transmission across \( N_c \) subcarriers. The OFDM system transmits \( N_m \leq min(N_r,N_t) \) data streams. For subcarrier $i$ and receiving antenna $j$, let $\mathbf{h}_{i,j} \in \mathbb{C}^{N_t\times1}$ denote the channel vector, $\mathbf{v}_{i,j} \in \mathbb{C}^{N_t\times N_m}$ denote transmit precoding vector, $x_{i,j}\in \mathbb{C}^{N_m\times 1}$ denote the transmitted data symbol, and $n_{i,j}\in \mathbb{C}$ denote the additive noise. Correspondingly,
the received signal of the UE at subcarrier $i$ and receiving antenna $j$ are given by:
\begin{equation}
{y}_{i,j} =\mathbf{\tilde{h}}_{i,j}^{H}\mathbf{v}_{i,j}x_{i,j}+{n}_{i,j}.
\end{equation}
 At the \( t \)-th transmission time step, the time-frequency response of the channel for all subcarriers and antennas is denoted by \( \mathbf{\tilde{H}}[t] = \{\mathbf{\tilde{h}}_{1,1}^{H}[t], \ldots, \mathbf{\tilde{h}}_{i,j}^{H}[t], \ldots, \mathbf{\tilde{h}}_{N_c,N_r}^{H}[t]\} \), where \( 1 \leq i \leq N_c \) and \( 1 \leq j \leq N_r \).

\subsection{Downlink CSI Prediction }
Downlink CSI prediction refers to the use of CSI data from the previous $l_m$ transmission time step to predict the CSI data for the next transmission time step. The modeling process is described as:
\begin{equation}
Train:\mathbf{\hat{H}}[t] = f(\mathbf{\tilde{H}}[t-1],\mathbf{\tilde{H}}[t-2],\dots ,\mathbf{\tilde{H}}[t-l_m],\tilde{\theta}_{M} ),
\end{equation}
\begin{equation}
Infer:\mathbf{\hat{H}}[t] = f(\mathbf{\tilde{H}}[t-1],\mathbf{\tilde{H}}[t-2],\dots ,\mathbf{\tilde{H}}[t-\tilde{l} ],\tilde{\theta}_{M}),
\end{equation}
where  $1\le \tilde{l}\le l_m$ represents the variable step length during the inference stage. $\tilde{\theta}_{M}$ refers to the learnable parameters of Csi-LLM, which are fine-tuned based on LLMs. It is important to note that in traditional downlink CSI prediction, $l_m$ is a hyperparameter that must be determined during the training stage, and the step length during the inference stage is also fixed at $l_m$. However, in Csi-LLM, $l_m$ represents the maximum step length in the collected channel data, and the length during the inference stage can be any value less than the maximum step length. This ensures maximum utilization of the training data and allows dynamic adjustment of the step length during the inference stage.

\begin{figure*}[htbp]
    \centering
    \includegraphics[width=0.8\linewidth,scale=1.00]{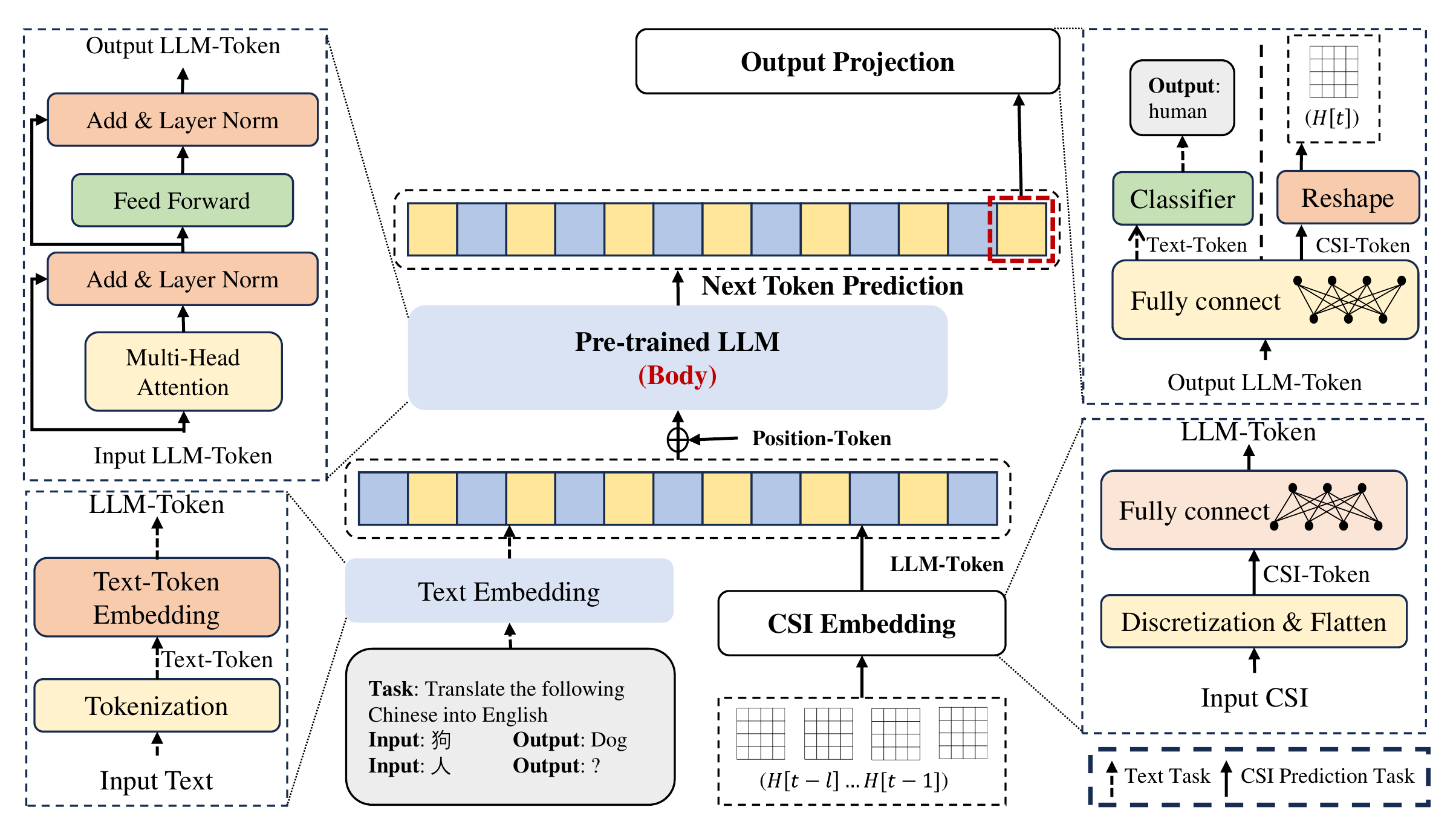}
    \caption{Csi-LLM network architecture overview.(The solid line represents the forward propagation of Csi-LLM, while the dashed line indicates the comparison for the text task.)}
    \label{fig-Network-architecture}
\end{figure*}

\section{Csi-LLM}

Large language foundation models are general models of language that are designed to support a large vairty of AI tasks\cite{dubey2024llama}. The development of modern foundation models consists of two main stages: \textbf{(1)} a pre-training stage in which the model is trained at massive scale using straightforward tasks such as next token generation and \textbf{(2)} a post-training stage in which the model is tuned to improve specific capabilities in downstream tasks. 

One of the key focuses in modern AI systems is how to leverage the general token sequence modeling capabilities acquired during the pre-training stage to enhance the performance of downstream tasks in the post-training stage. Alignment is considered one of the most effective methods to achieve this\cite{li2010survey}. To maximize the utilization of LLMs' pattern recognition and reasoning abilities, Csi-LLM aligns the training process of LLMs in three aspects: network structure, data processing, and optimization objectives.

\subsection{Network Architecture Alignment}

The LLM network architecture consists of three components: Embedding, Pre-trained LLM, and Classifier. \textbf{Embedding}: This component embeds input data from different modalities into a unified representation known as LLM-Token. The LLM-Token is the smallest atomic input for the Pre-trained LLM. \textbf{Pre-trained LLM}: Based on the Transformer deep neural network architecture, this component undergoes unsupervised training on a vast corpus of text data. It possesses a powerful next token generation capability. \textbf{Classifier}: This component performs post-processing on the generated next token and outputs the final modality data. As shown in Fig. \ref{fig-Network-architecture}, Csi-LLM aligns with the existing LLM network structure and is divided into three parts: \textbf{CSI-Embedding}, \textbf{Pre-trained LLM}, and \textbf{Output Projection}.

As shown in the upper left part of Fig. \ref{fig-Network-architecture}, Csi-LLM has opted for a causal decoder architecture for the Pre-trained LLM based on unidirectional attention. LLMs typically follow three mainstream architectures: causal decoder, prefix decoder, and encoder-decoder\cite{zhao2023survey}. The causal decoder architecture uses a unidirectional attention mask, ensuring that each input LLM-token can only attend to past tokens and itself. This unidirectional attention mechanism is highly suitable for temporal prediction, as temporal prediction can only consider historical temporal information and cannot focus on future temporal information.

Specifically, we chose GPT-2\cite{radford2019language} as the core component of Csi-LLM. Compared to LLMs with billions of parameters, the earlier version developed by OpenAI, GPT-2, with its millions of parameters, is more suitable for early-stage research on LLM capabilities in wireless communication scenarios. More importantly, researchers have found that GPT-2-based time series models exhibit highly competitive performance across various temporal tasks, such as network traffic prediction, weather forecasting, and stock price prediction\cite{zhou2023one}.

Csi-LLM differs from the traditional neural network forward propagation process, aligning instead with the generation method of existing LLMs. In traditional neural network architectures, all neurons in the intermediate layers participate in the computation of subsequent modules. However, in LLMs, only the last token is involved in the final Classifier module. As illustrated in Fig. \ref{fig-Network-architecture}, in Csi-LLM, among the neurons output by the Pre-trained LLM, only the last LLM-Token is passed forward to the Output Projection for the final calculation of CSI data.

After aligning with the LLM network structure, Csi-LLM can handle variable-length input data. Since only the last LLM-Token participates in the Output Projection calculation process, the structure design of Output Projection is immune to the number of LLM-Tokens in the shallow network. This enables Csi-LLM to support historical sequences of any length as input and output the next step of CSI prediction results.

\subsection{Data Processing Alignment}
In Csi-LLM, the CSI-Embedding and Output Projection are designed to achieve the conversion between CSI data and the LLM-Token representation. The embedding process typically involves data discretization and sequence encoding. As shown in the lower left part of Fig. \ref{fig-Network-architecture}, in Text Embedding, a long text input is first tokenized into Text-Token (usually characters or phrases), which are then encoded into a numerical vector LLM-Token through a fully connected layer (the encoding strategy of GPT-2). As shown in the lower right part of Fig. \ref{fig-Network-architecture}, CSI-Embedding aligns with the embedding method of GPT-2. In this process, tokenization directly discretizes the data by time steps, with each step of CSI data corresponding to a Csi-Token. The multi-dimensional Csi-Token data is then flattened into one-dimensional linear data, and, similar to GPT-2, a fully connected layer is used to achieve compressed encoding of the data.

The parallel computation of the Transformer structure relies on additional Position-Tokens to provide positional information. GPT-2 employs a learnable positional encoding strategy, which is internalized into the LLM foundation model during pre-training. Therefore, Csi-LLM uses the same Position-Tokens as in the pre-training stage.

In the Output Projection module, Csi-LLM utilizes a nonlinear fully connected layer to map LLM-Tokens to CSI data representation. Typically, LLMs perform discriminative tasks in Output Projection using fully connected layers, as the LLM-Tokens in text data form a finite, enumerable set with a limited number of characters. However, CSI data represents continuous, non-enumerable physical signals. Therefore, Csi-LLM employs a nonlinear fully connected layer in Output Projection to directly represent the data.

\subsection{Optimization objectives Alignment}
To maximize the use of LLM capabilities for CSI prediction tasks, the optimization objective of Csi-LLM must align with the optimization objective of LLM pre-training. Fortunately, since most language tasks can be cast as the prediction problem based on the input, the language modeling tasks used in LLM pre-training share a similar learning process with temporal prediction tasks.

The language modeling task (LM) is the most commonly used objective for pre-training causal decoder LLMs, such as GPT-2\cite{radford2019language} and PaLM\cite{JMLR:v24:22-1144}. Given a sequence of tokens $\textbf{T}={T_1,...,T_n}$, the LM task aims to autoregressively predict the next target tokens $T_i$ based on the preceding tokens $T_{<i}$ in a sequence. 

Similarly, the channel prediction task addressed by Csi-LLM involves given CSI data over a period $[{\mathbf{\tilde{H}}[1],...,\mathbf{\tilde{H}}[t]}]$. The objective is to autoregressively predict the CSI 
at the next time step $\mathbf{\tilde{H}}[i]$ based on the preceding CSI $\mathbf{\tilde{H}}[<i]$. Therefore, Csi-LLM departs from the traditional approach of using fixed-length historical CSI as input and single-step future CSI as output for optimization. Instead, it aligns with the next token generation process in LLM pre-training, which we refer to as next step CSI prediction. The training loss function can be expressed as:

\begin{equation}
loss =\sum_{i=0}^{l_m-1}  \text{MSE}(\mathbf{\tilde{H}}(t-i),\mathbf{\hat{H}}(t-i)),
\end{equation}

where $\mathbf{\tilde{H}}$ and $\mathbf{\hat{H}}$ denote the actual and predicted next step CSI, respectively.

\section{Experiment}
\subsection{Experimental Settings}
\subsubsection{Datasets}
The dataset used in this study comes from an open mobile communication dataset\footnote{www.mobileai-dataset.com}. It contains a total of four data files, corresponding to four speed scenarios: 30 km/h, 60 km/h, 120 km/h, and a mix of the three speeds. Each scenario dataset includes 21,000 samples, encompassing time-domain information over 20 transmission time steps and frequency-domain channel information for 8 physical resource blocks (PRBs), with a transmission time interval of 5 ms. The specific parameter settings are shown in Table \ref{setting}.

\subsubsection{Performance Indices}
We use normalized mean squared error(NMSE) as the performance indices, which are defined as follows:
\begin{equation}
    \text{NMSE}(\mathbf{\tilde{H}}(t),\mathbf{\hat{H}}(t))  =  \frac{1}{N} \sum_{i = 1}^{N}\frac{\left \|  \mathbf{\tilde{H}}(t)-\mathbf{\hat{H}}(t)  \right \| _2 }{\left \|  \mathbf{\tilde{H}}(t)  \right \| _2 }  ,
\end{equation}
where $\mathbf{\tilde{H}}$, $\mathbf{\hat{H}}$ represent the actual and predicted downlink CSI.

\begin{table}[htbp]
\caption{Dataset Parameter Settings}
\centering
\label{setting}
\setlength{\tabcolsep}{2mm}
{

\begin{tabular}{c|c}
\toprule
\textbf{Parameters}                 & \textbf{Value}  \\ \midrule
\textbf{Scenario}          & Dense Urban(Macro only) \\
\textbf{Inter-BS distance} & 200m                    \\
\textbf{Channel model}     & According to TR 38.901  \\
\textbf{Frequency Range}   & FR1 only; 2GHz          \\
\textbf{Bandwidth}         & 10M(52RB)               \\
\textbf{SCS}               & 15kHz for 2GHz          \\
\textbf{Data size}         & (21000,20,2,32,4,8) \\
\textbf{Speed}             & 30/60/120/Mix. km/h    \\
\textbf{Number of transmit antennas ($N_t$)}    & 32              \\
\textbf{Number of receiver antennas ($N_r$)}    & 4              \\
\bottomrule
\end{tabular}
}
\end{table}

\begin{table}[htbp]
\centering
\caption{The Training-Related Parameter Settings}
\begin{tabular}{cccccc}
\toprule
\textbf{Parmeters(GPT-2)} & \textbf{Layers} & \textbf{$d_{model}$} & \textbf{$l_m$} & \textbf{Batch Size}  & \textbf{lr}   \\ \midrule
117M      & 12     & 768      & 16   & 512 & 1e-3 \\
\bottomrule
\end{tabular}
\label{table2}
\end{table}

\subsubsection{Training Setup}
GPT-2 includes four different size versions \cite{radford2019language}, and we selected the smallest model for our experiments. Due to the dataset being limited to 20 transmission time steps, the maximum step length $l_m$ for training was set to 16. The other training-related parameters are shown in Table \ref{table2}. The numbers of samples in the training, validation, and test sets are 17,640, 1,680, and 1,680, respectively.

\begin{table*}[]
\centering
\caption{ 
One-step Prediction Results of Traditional Solutions and Csi-LLM in Different Scenarios.The best performances are highlighted in \textbf{bold}, and suboptimal ones are marked with an \underline{underline}.}
\label{tabel3}
\scalebox{1}
{
\begin{tabular}{c|c|c|c|c|c|c}
\toprule
\multirow{2}{*}{\textbf{Speed}} & \multirow{2}{*}{\textbf{Input Step Length}} & \textbf{No-prediction} & \textbf{Fixed-step-size-4} & \textbf{Fixed-step-size-8} & \textbf{Fixed-step-size-16} & \textbf{Csi-LLM}  \\
                                &                                             & \textbf{NMSE}          & \textbf{NMSE}         & \textbf{NMSE}         & \textbf{NMSE}          & \textbf{NMSE}     \\
\midrule
\multirow{4}{*}{30 km/h}             & 2                                           & \underline{ -6.7453}          & -                     & -                     & -                      & \textbf{-14.6400} \\
                                & 4                                           & -6.7207                & \underline{ -14.1759}        & -                     & -                      & \textbf{-15.2657} \\
                                & 8                                           & -6.7209                &     -           & \underline{-13.7007 }             & -                      & \textbf{-15.2986} \\
                                & 16                                          & -6.7310                &      -          &       -                & \underline{-13.5928  }             & \textbf{-15.2631} \\
\midrule
\multirow{4}{*}{60 km/h}             & 2                                           & \underline{ -3.1634}          & -                     & -                     & -                      & \textbf{-13.3911} \\
                                & 4                                           & -3.1647                & \underline{ -13.2507}        & -                     & -                      & \textbf{-13.5297} \\
                                & 8                                           & -3.1584                &   -             & \underline{-8.3404 }              & -                      & \textbf{-13.5387} \\
                                & 16                                          & -3.1526                &   -             &       -                & \underline{-8.0612  }              & \textbf{-13.5583} \\
\midrule
\multirow{4}{*}{120 km/h}            & 2                                           & \underline{ -5.5650}          & -                     & -                     & -                      & \textbf{-13.4626} \\
                                & 4                                           & -5.5723                & \underline{ -13.1522}        & -                     & -                      & \textbf{-13.4925} \\
                                & 8                                           & -5.6081                &      -          & \underline{-12.9521   }           & -                      & \textbf{-13.5933} \\
                                & 16                                          & -5.6028                &    -            &       -                &\underline{ -7.7057   }             & \textbf{-13.6053} \\
\midrule
\multirow{4}{*}{Mixture Speed}            & 2                                           & \underline{ -4.7779}          & -                     & -                     & -                      & \textbf{-13.2791} \\
                                & 4                                           & -4.7872                & \underline{ -13.1080}        & -                     & -                      & \textbf{-13.4215} \\
                                & 8                                           & -4.7656                &  -              &\underline{ -7.8414}               & -                      & \textbf{-13.4651} \\
                                & 16                                          & -4.7798                &   -             &       -                & \underline{-7.5567   }             & \textbf{-13.4430}

\\
\bottomrule
\end{tabular}
}
\end{table*}

\subsection{Prediction Performance}

\subsubsection{One-step Prediction}

In this subsection, we report the prediction accuracy of Csi-LLM and two traditional channel prediction schemes. The traditional prediction schemes are as follows:
\begin{itemize}
    \item No-prediction: The downlink CSI of the last time step is used as the downlink CSI of the next time step, which illustrates the phenomenon of channel aging.
    \item Fixed-step-size-4/8/16\cite{10440484}: The downlink CSI from the past 4, 8, or 16 steps is used as the neural network input to predict the downlink CSI of the next time step.
\end{itemize}
To ensure the fairness of the experiments, Fixed-step-size-4/8/16 maintains the same architecture as Csi-LLM in terms of the embedding of raw data and the Transformer-based network body. However, its output projection module and optimization method do not align with the pre-training of LLMs but instead follow traditional neural network channel prediction schemes. The main differences from the aligned Csi-LLM are reflected in two aspects:

Firstly, as described in Section \uppercase\expandafter{\romannumeral3}.A, traditional channel prediction network schemes fix the mapping between input and output, requiring all intermediate layer neurons' outputs to participate in subsequent module computations. Therefore, different input step lengths necessitate different network structures for output projection. As a result, the output projection module of Fixed-step-size-4/8/16 must be adjusted according to changes in input step length and retrained end-to-end. In contrast, Csi-LLM involves only the last token in the output projection module's computation process, making its output projection module design independent of the input step length, thereby supporting any step length as input.

Secondly, as mentioned in Section \uppercase\expandafter{\romannumeral3}.C, traditional channel prediction network schemes follow a typical supervised learning process, where a fixed step length of historical downlink CSI is input, and the next time step's downlink CSI serves as the label. In contrast, Csi-LLM undergoes unsupervised training under autoregressive prediction.

Table \ref{tabel3} presents the prediction results of Csi-LLM and the two traditional schemes for input step lengths of 2, 4, 8, and 16 in four speed scenarios. Due to the fixed relationship between input and output in Fixed-step-size-4/8/16, it can only perform channel prediction under one specific condition and fails in other scenarios. In contrast, Csi-LLM demonstrates stable and optimal performance in all scenarios, indicating that Csi-LLM has superior scalability and practicality compared to traditional schemes.

\subsubsection{Continuous Autoregressive Prediction}
In this subsection, we continue to report the performance of Csi-LLM and other traditional schemes in continuous autoregressive prediction of multiple time steps for downlink CSI. As described in Section \uppercase\expandafter{\romannumeral1}, to prevent the rapid accumulation of prediction errors during successive sequential channel prediction, previous work has proposed multi-step parallel prediction. Therefore, building on the previous section, we include existing multi-step parallel prediction as a comparative scheme. The traditional parallel prediction schemes are as follows:
\begin{itemize}
    \item Fixed-step-size-4-parallel\cite{jiang2022accurate}: The downlink CSI from the past 4 steps is used as the neural network input to parallel predict the downlink CSI of the next 4 time steps.
\end{itemize}

In this experiment, the variable input length of Csi-LLM allows historical CSI to always be retained in the input window during sequential prediction. In contrast, traditional Fixed-step-size-4 and Fixed-step-size-4-parallel schemes must have an input window of 4 steps, thus gradually removing more distant historical CSI during sequential prediction.

It is important to note that, compared to the optimization approach of traditional channel prediction schemes, the optimization direction of Csi-LLM described in \uppercase\expandafter{\romannumeral3}.C is more suited to practical continuous channel prediction scenarios.


\begin{figure}[htbp]
    \centering
    \includegraphics[width=\linewidth]{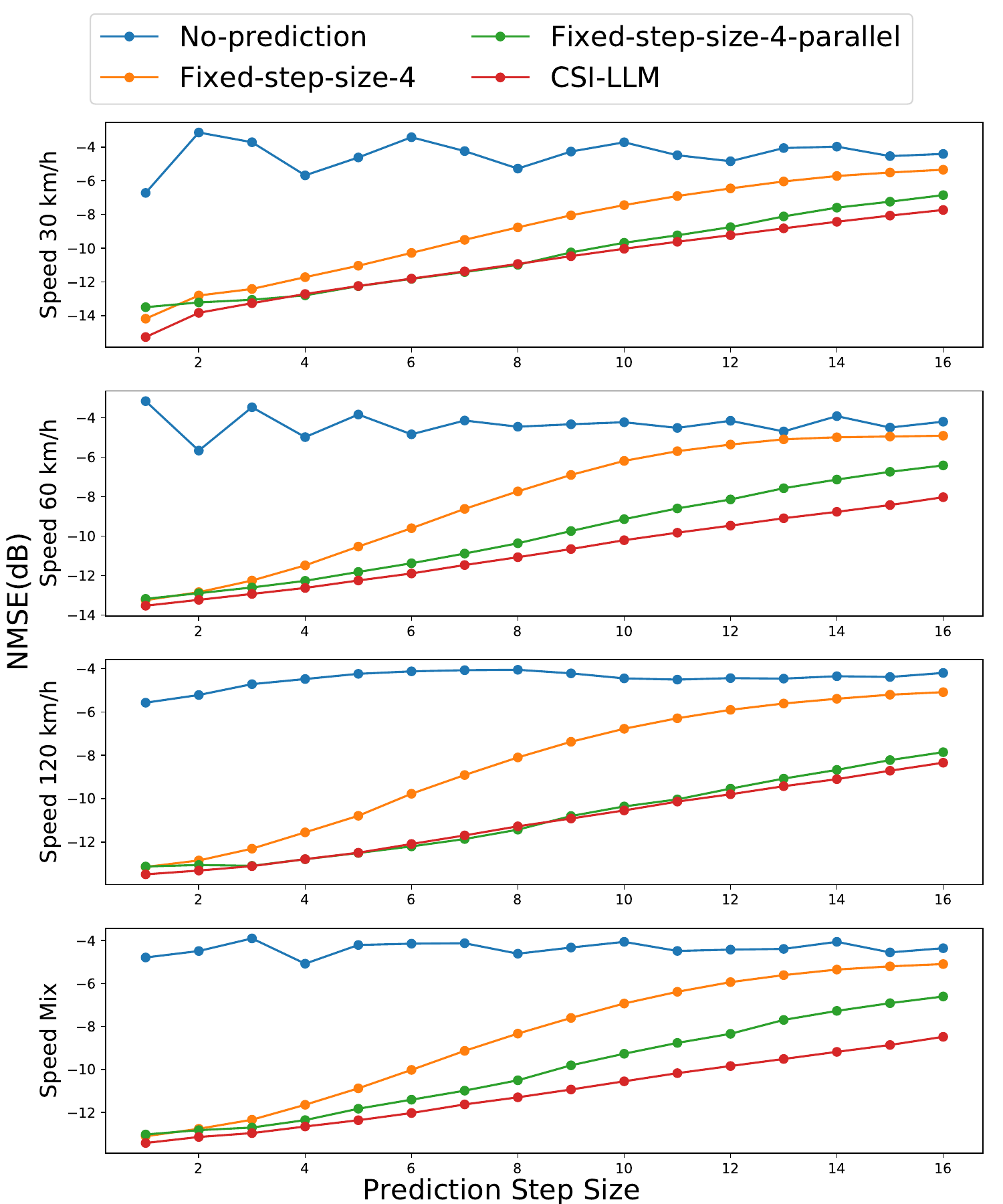}
    \caption{The comparative experimental results of continuous autoregressive downlink channel prediction.}
    \label{fig3}
\end{figure}

As shown in Fig. \ref{fig3}, the continuous autoregressive channel prediction results of Csi-LLM and the three traditional channel prediction schemes over the next 16 time steps in four different speed scenarios. The experimental results indicate that Csi-LLM exhibits better performance in mitigating performance degradation.

\begin{figure}[htbp]
    \centering
    \includegraphics[width=\linewidth]{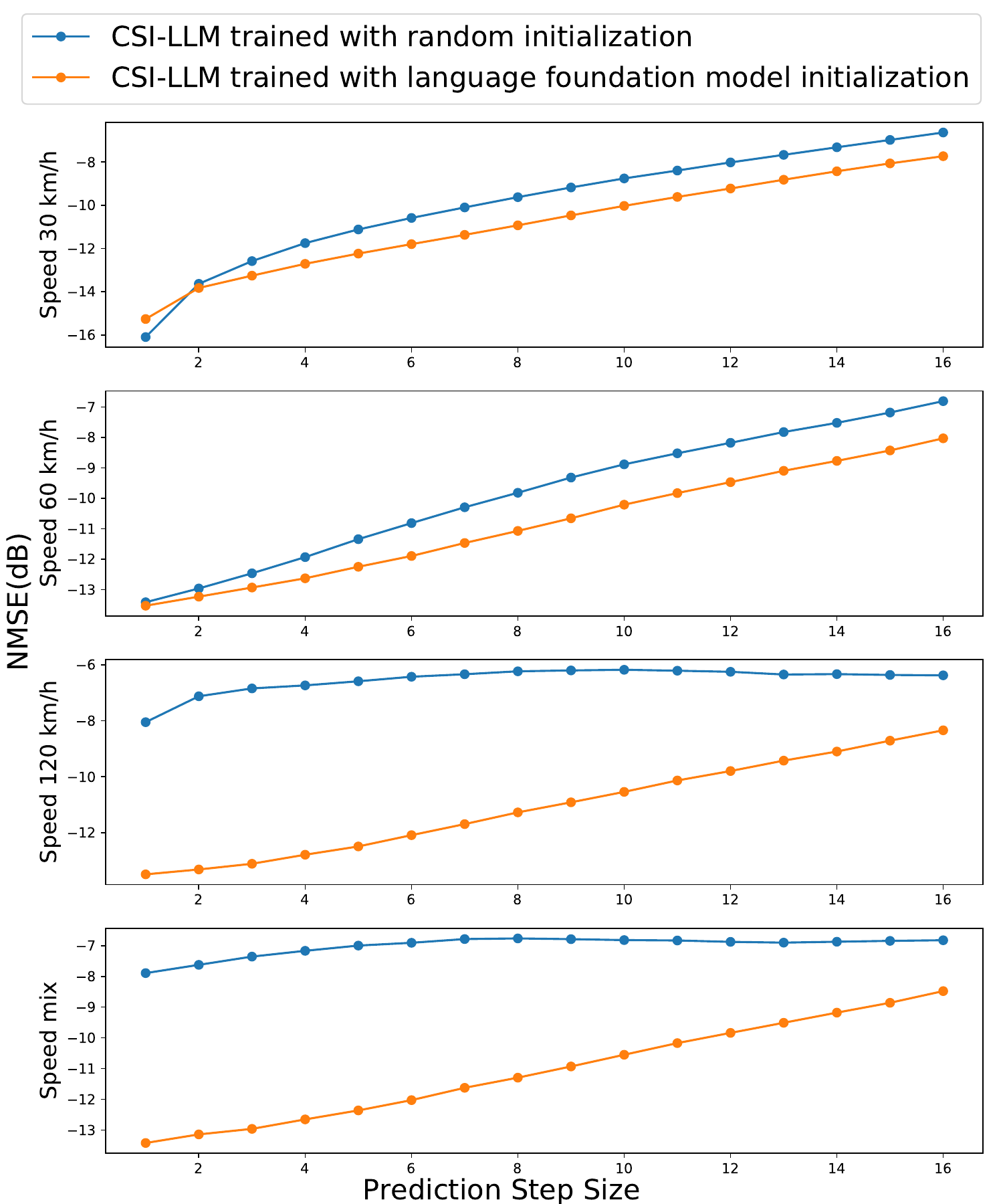}
    \caption{The impact of language foundation models on the downlink channel prediction performance of Csi-LLM.}
    \label{fig4}
\end{figure}

\subsection{Ablation experiment}
A fundamental difference between Csi-LLM and traditional channel prediction schemes lies in the introduction of a foundation model pre-trained on a large corpus of text. To further investigate the impact of text pre-training tasks on downlink channel prediction tasks, we compare the performance of Csi-LLM initialized randomly versus initialized from the foundation model.

As shown in Fig. \ref{fig4}, Csi-LLM derived from the foundation model exhibits more accurate downlink channel prediction performance, especially in high-speed (120 km/h) and complex mixed-speed scenarios. This indicates that the token modeling capabilities acquired during the text pre-training task can be transferred to the wireless communication domain. Additionally, the complex pattern recognition abilities learned by the foundation model in text tasks help optimize more intricate downstream tasks, enhancing the model's ability to relearn and improve.

\section{Conclusion}
To design a flexible model that adapts to varying mobility scenarios, we developed Csi-LLM using the GPT-2 model. Compared to traditional downlink channel prediction methods, Csi-LLM not only achieves higher prediction accuracy but also demonstrates enhanced error resilience in continuous sequential prediction. Furthermore, Csi-LLM’s design is aligned with the ongoing advancements in large language models, allowing it to leverage future improvements in LLM architectures and training methodologies. This alignment ensures that Csi-LLM remains a robust and scalable solution for downlink channel prediction, effectively adapting to the evolving demands of wireless communication systems. The results underscore the potential of integrating LLMs into wireless communication tasks, paving the way for new research and application opportunities in this field.



\bibliographystyle{IEEEtran}
\bibliography{ref}

\begin{thebibliography}{10}
\providecommand{\url}[1]{#1}
\csname url@samestyle\endcsname
\providecommand{\newblock}{\relax}
\providecommand{\bibinfo}[2]{#2}
\providecommand{\BIBentrySTDinterwordspacing}{\spaceskip=0pt\relax}
\providecommand{\BIBentryALTinterwordstretchfactor}{4}
\providecommand{\BIBentryALTinterwordspacing}{\spaceskip=\fontdimen2\font plus
\BIBentryALTinterwordstretchfactor\fontdimen3\font minus \fontdimen4\font\relax}
\providecommand{\BIBforeignlanguage}[2]{{%
\expandafter\ifx\csname l@#1\endcsname\relax
\typeout{** WARNING: IEEEtran.bst: No hyphenation pattern has been}%
\typeout{** loaded for the language `#1'. Using the pattern for}%
\typeout{** the default language instead.}%
\else
\language=\csname l@#1\endcsname
\fi
#2}}
\providecommand{\BIBdecl}{\relax}
\BIBdecl

\bibitem{truong2013effects}
K.~T. Truong and R.~W. Heath, ``Effects of channel aging in massive mimo systems,'' \emph{Journal of Communications and Networks}, vol.~15, no.~4, pp. 338--351, 2013.

\bibitem{duel2007fading}
A.~Duel-Hallen, ``Fading channel prediction for mobile radio adaptive transmission systems,'' \emph{Proceedings of the IEEE}, vol.~95, no.~12, pp. 2299--2313, 2007.

\bibitem{baddour2005autoregressive}
K.~E. Baddour and N.~C. Beaulieu, ``Autoregressive modeling for fading channel simulation,'' \emph{IEEE Transactions on Wireless Communications}, vol.~4, no.~4, pp. 1650--1662, 2005.

\bibitem{yin2020addressing}
H.~Yin, H.~Wang, Y.~Liu, and D.~Gesbert, ``Addressing the curse of mobility in massive mimo with prony-based angular-delay domain channel predictions,'' \emph{IEEE Journal on Selected Areas in Communications}, vol.~38, no.~12, pp. 2903--2917, 2020.

\bibitem{kim2020massive}
H.~Kim, S.~Kim, H.~Lee, C.~Jang, Y.~Choi, and J.~Choi, ``Massive mimo channel prediction: Kalman filtering vs. machine learning,'' \emph{IEEE Transactions on Communications}, vol.~69, no.~1, pp. 518--528, 2020.

\bibitem{jiang2019neural}
W.~Jiang and H.~D. Schotten, ``Neural network-based fading channel prediction: A comprehensive overview,'' \emph{IEEE Access}, vol.~7, pp. 118\,112--118\,124, 2019.

\bibitem{jiang2022accurate}
H.~Jiang, M.~Cui, D.~W.~K. Ng, and L.~Dai, ``Accurate channel prediction based on transformer: Making mobility negligible,'' \emph{IEEE Journal on Selected Areas in Communications}, vol.~40, no.~9, pp. 2717--2732, 2022.

\bibitem{openai2024gpt4technicalreport}
\BIBentryALTinterwordspacing
OpenAI, J.~Achiam, S.~Adler, S.~Agarwal, L.~Ahmad, I.~Akkaya, F.~L. Aleman, D.~Almeida, and J.~Altenschmidt, ``Gpt-4 technical report,'' 2024. [Online]. Available: \url{https://arxiv.org/abs/2303.08774}
\BIBentrySTDinterwordspacing

\bibitem{JMLR:v24:22-1144}
\BIBentryALTinterwordspacing
A.~Chowdhery, S.~Narang, J.~Devlin, M.~Bosma, G.~Mishra, A.~Roberts, P.~Barham, H.~W. Chung, C.~Sutton, and S.~Gehrmann, ``Palm: Scaling language modeling with pathways,'' \emph{Journal of Machine Learning Research}, vol.~24, no. 240, pp. 1--113, 2023. [Online]. Available: \url{http://jmlr.org/papers/v24/22-1144.html}
\BIBentrySTDinterwordspacing

\bibitem{tong2023ten}
W.~Tong, C.~Peng, T.~Yang, F.~Wang, J.~Deng, R.~Li, L.~Yang, H.~Zhang, D.~Wang, M.~Ai \emph{et~al.}, ``Ten issues of netgpt,'' \emph{arXiv preprint arXiv:2311.13106}, 2023.

\bibitem{9839111}
X.~Liang, J.~Shen, H.~Chang, X.~Gu, and L.~Zhang, ``Deep learning-based cooperative csi feedback via multiple receiving antennas in massive mimo,'' in \emph{ICC 2022 - IEEE International Conference on Communications}, 2022, pp. 1373--1378.

\bibitem{dubey2024llama}
A.~Dubey, A.~Jauhri, A.~Pandey, A.~Kadian, A.~Al-Dahle, A.~Letman, A.~Mathur, A.~Schelten, A.~Yang, A.~Fan \emph{et~al.}, ``The llama 3 herd of models,'' \emph{arXiv preprint arXiv:2407.21783}, 2024.

\bibitem{li2010survey}
H.~Li and N.~Homer, ``A survey of sequence alignment algorithms for next-generation sequencing,'' \emph{Briefings in bioinformatics}, vol.~11, no.~5, pp. 473--483, 2010.

\bibitem{zhao2023survey}
W.~X. Zhao, K.~Zhou, J.~Li, T.~Tang, X.~Wang, Y.~Hou, Y.~Min, B.~Zhang, J.~Zhang, Z.~Dong \emph{et~al.}, ``A survey of large language models,'' \emph{arXiv preprint arXiv:2303.18223}, 2023.

\bibitem{radford2019language}
A.~Radford, J.~Wu, R.~Child, D.~Luan, D.~Amodei, I.~Sutskever \emph{et~al.}, ``Language models are unsupervised multitask learners,'' \emph{OpenAI blog}, vol.~1, no.~8, p.~9, 2019.

\bibitem{zhou2023one}
T.~Zhou, P.~Niu, L.~Sun, R.~Jin \emph{et~al.}, ``One fits all: Power general time series analysis by pretrained lm,'' \emph{Advances in neural information processing systems}, vol.~36, pp. 43\,322--43\,355, 2023.

\bibitem{10440484}
S.~Fan, H.~Li, X.~Liang, Z.~Liu, X.~Gu, and L.~Zhang, ``E2enet: An end-to-end channel prediction neural network based on uplink pilot for fdd systems,'' \emph{IEEE Wireless Communications Letters}, vol.~13, no.~5, pp. 1285--1289, 2024.

\end{thebibliography}

\end{document}